\documentclass[superscriptaddress, twocolumn,showpacs,
amssymb,amsmath,nobibnotes,aps,longbibliography,
nofootinbib]{revtex4-2}
\usepackage{graphicx,subfigure,bm,color,psfrag,hyperref}
\usepackage{amsfonts}
\usepackage{lipsum}
\usepackage{mathtools}
\usepackage{verbatim}
\usepackage[normalem]{ulem}
\usepackage[dvipsnames]{xcolor}
\hypersetup{colorlinks,linkcolor={blue},citecolor={red},urlcolor={cyan}}

\begin{document}

\begin{center}    
\end{center}
\title{Topological and optical signatures of modified black-hole entropies}
\begin{center}   
\end{center}

\author{Ankit Anand}
\email{anand@iitk.ac.in} 
\affiliation{Department of Physics, Indian Institute of Technology, Kanpur 
208016, India}

\author{Kimet Jusufi}
\email{kimet.jusufi@unite.edu.mk}
\affiliation{Physics Department, State University of Tetovo, Ilinden Street nn, 
1200, Tetovo, North Macedonia.}

\author{Spyros~Basilakos}
\email{svasil@academyofathens.gr}
\affiliation{\mbox{National Observatory of Athens, Lofos Nymfon, 11852 Athens, 
Greece}}
\affiliation{Academy of Athens, Research Center for Astronomy and Applied 
Mathematics, Soranou Efesiou 4, 11527, Athens, Greece}
\affiliation{School of Sciences, European University Cyprus, Diogenes Street, 
Engomi, 1516 Nicosia, Cyprus}

 \author{Emmanuel N. Saridakis} 
\email{msaridak@noa.gr}
\affiliation{National Observatory of Athens, Lofos Nymfon, 11852 Athens, Greece}
\affiliation{Departamento de Matem\'{a}ticas, Universidad Cat\'{o}lica del 
Norte, Avda. Angamos 0610, Casilla 1280 Antofagasta, Chile}
\affiliation{CAS Key Laboratory for Researches in Galaxies and Cosmology, 
School 
of Astronomy and Space Science, University of Science and Technology of China, 
Hefei, Anhui 230026, China}


\begin{abstract}
We investigate how deviations from the Bekenstein-Hawking entropy modify 
black-hole spacetimes through the recently proposed entropy-geometry 
correspondence. For four representative modified entropies, namely Barrow, 
R\'enyi, 
Kaniadakis, and logarithmic, we derive the corresponding effective  
metrics 
and analyze their thermodynamic and topological classification using the 
off-shell free energy and 
winding numbers. We show that Barrow and R\'enyi entropies yield a single 
unstable sector with 
global charge $W=-1$, while logarithmic and Kaniadakis corrections produce 
canceling defects with $W=0$, revealing topological structures absent in the 
Schwarzschild case. 
Using the modified metrics, we further calculate the photon-sphere radius and 
shadow size, showing that each modified entropy relation induces characteristic 
optical 
shifts. Thus, by comparing with Event Horizon Telescope observations of 
Sgr~A$^\ast$, we extract  new bounds on all entropy-deformation 
parameters. Our results 
demonstrate that thermodynamic topology, together with photon-sphere 
phenomenology, offers a viable way to test generalized entropy 
frameworks 
and probe departures from the Bekenstein-Hawking area law.

\end{abstract}

\maketitle

\section{Introduction}

The known connection between gravity, thermodynamics, and quantum theory has 
long been recognized as essential for understanding the structure of spacetime. 
Since the pioneering works of Bekenstein and Hawking that  established black 
holes as thermodynamic systems possessing temperature and entropy 
\cite{Bekenstein:1973ur,Hawking:1975vcx}, it has become clear that 
gravity exhibits an intrinsically thermodynamic character. The area law and the 
holographic principle further suggest that spacetime may contain microscopic 
degrees of freedom, with classical gravitational dynamics emerging as a 
macroscopic, statistical description 
\cite{tHooft:1993dmi,Susskind:1994vu,Padmanabhan:2009vy,Verlinde:2010hp}.

Corrections to the Bekenstein-Hawking entropy are therefore of central 
importance in extending semiclassical gravity. Although the leading 
contribution 
$S=A/4$ arises from quantum fields near the horizon, a wide range of approaches 
predict subleading logarithmic, power-law,  or exponential 
contributions. In particular, one can consider    
non-extensive entropic corrections  such as in Tsallis 
\cite{Tsallis:1987eu,volovik2024tsalliscirto},  R\'enyi 
\cite{Renyi:1961,aa,bb}
and and 
Sharma-Mittal~\cite{Sharma1975} entropies, quantum-gravitational corrections  
such as 
in Barrow entropy \cite{Barrow:2020tzx}, relativistic corrections such as in 
Kaniadakis entropy  \cite{PhysRevE.66.056125,PhysRevE.72.036108},   logarithmic 
 corrections \cite{Kaul:2000kf,Das:2001ic, Sen:2012dw}, etc.
Such 
generalized entropy frameworks, substantially modify both the thermodynamic 
relations and the associated geometrical properties, leading to a rich 
phenomenology in black-hole and 
cosmological contexts 
\cite{Lymperis:2018iuz,Saridakis:2020lrg,Moradpour:2018ivi,Nojiri:2019skr,
Abreu:2017hiy, Iqbal:2019ooy,
Maity:2019qbv,Geng:2019shx,Lymperis:2021qty,Mohammadi:2021wde,
Telali:2021jju,Hernandez-Almada:2021rjs,Jusufi:2022mir,Komatsu:2022bik,
Zamora:2022cqz,Drepanou:2021jiv,Luciano:2022ely,Jusufi:2021fek,Nojiri:2022dkr,
Luciano:2022knb,
Jizba:2022bfz,Chanda:2022tpk,
Dheepika:2022sio,Saha:2022oph,
Luciano:2023zrx,
Luciano:2023fyr,Teimoori:2023hpv,Naeem:2023ipg,Jalalzadeh:2023mzw,
Basilakos:2023kvk, Naeem:2023tcu, Coker:2023yxr, Lymperis:2023prf,
Saavedra:2023rfq,Nakarachinda:2023jko,Jizba:2023fkp,
Okcu:2024tnw,Jalalzadeh:2024qej,Adhikary:2024sax,Jawad:2024yrv,Sheykhi:2024fya, 
Jizba:2024klq,Huang:2024xqk, 
Ebrahimi:2024zrk,
Trivedi:2024inb, Yarahmadi:2024oqv,
 Karabat:2024trf,Ens:2024zzs,Tsilioukas:2024seh,
Ualikhanova:2024xxe,Shahhoseini:2025sgl,Lymperis:2025vup,Nojiri:2025gkq,
Luciano:2025elo,Luciano:2025hjn,Jusufi:2024rba,Dabrowski:2024qkp,Jusufi:2023ayv,
Jusufi:2025hte,Basilakos:2025wwu,Kotal:2025bof,
Nojiri:2025fiu,Capozziello:2025axh,Abreu:2025etn,Luciano:2025fqg}.

Motivated by these developments, in the present analysis we examine how 
logarithmic \cite{Kaul:2000kf,Sen:2013ns,Carlip:2000nv} and exponential 
\cite{Medved:2004tp,Nojiri:2005sr,Pourhassan:2017qrq,Chatterjee:2020iuf} 
entropy corrections deform spherically symmetric black-hole geometries.
In particular, within the emergent-gravity paradigm 
\cite{Jacobson:1995ab,Padmanabhan:2009vy}, 
gravity can be viewed as an entropic force generated by gradients of horizon 
entropy \cite{Verlinde:2010hp,Callen:1985}. From this perspective, any 
modification of the entropy-area relation induces corresponding deviations in 
the underlying spacetime metric. Very recently, in \cite{Anand:2025rjg} it was 
demonstrated that starting from a chosen entropy functional one may 
systematically derive the associated modified metric and its effective matter 
sector, establishing a direct entropy-geometry correspondence. This contrasts 
with earlier approaches \cite{Nojiri:2022ljp,Elizalde:2025iku}, in which 
generalized entropies were examined on pre-assigned geometrical backgrounds. 
The 
effective matter interpretation provided by the entropy deformation enables a 
physically coherent description of how quantum-gravitational or statistical 
corrections manifest as geometric backreaction.

In this work, we use these backreacted metrics to investigate two complementary 
types of signatures: thermodynamic topology and photon-sphere geometry. Recent 
studies \cite{Wei:2022dzw,Wei:2024gfz} have shown that black holes may be 
characterized not only by standard thermodynamic quantities, but also by global 
topological properties in parameter space. These appear as thermodynamic 
topological defects whose classification follows from winding numbers of the 
generalized free energy.  

In parallel, significant progress has been made in understanding photon spheres 
and black-hole shadows from a geometric standpoint. For static, spherically 
symmetric spacetimes, the photon sphere can be characterized in terms of 
vanishing geodesic curvature and Gaussian optical curvature, 
providing a coordinate-invariant framework for studying null circular orbits. 
Deviations from the standard Bekenstein-Hawking entropy, and thus from 
the Schwarzschild geometry, naturally shift both the photon-sphere radius and 
the shadow size. Hence, the latter can be directly confronted with 
observational 
measurements from the Event Horizon Telescope (EHT).  

In the analysis of the present manuscript we show how Barrow, R\'enyi, 
Kaniadakis, and logarithmic extended entropy relations  modify both the 
thermodynamic topology and the photon-sphere geometry of 
the associated black-hole solutions. Them using the EHT constraints on the 
shadow of 
Sgr A*, we derive observational bounds on each corresponding entropy parameter.

The paper is organized as follows.
In Section~\ref{sec:topo-defects}, we review the thermodynamic-topology 
framework and the interpretation of winding numbers.
In Section~\ref{Sec:TopologicalClassificationGeneral}, we present the 
general entropy-based topological classification and derive the photon-sphere 
conditions for arbitrary entropy deformations.
In Section~\ref{SecIV}, we apply this formalism to Barrow, R\'enyi, Kaniadakis, 
and logarithmic entropy expressions, computing their winding numbers, 
photon-sphere 
properties, and EHT-based parameter constraints.
Finally, Section~\ref{Sec:Conclusions} summarizes our results.

 \section{Thermodynamic topology and black holes as topological defects}
\label{sec:topo-defects}

In this section we summarise the thermodynamic-topological formalism
that associates black-hole equilibrium points with topological defects
in a two-dimensional parameter space $(r_h,\theta)$.
The presentation follows \cite{Wei:2022dzw,Wei:2024gfz},
and we enhance it with brief physical explanations so that both the
geometric and thermodynamic interpretations of each step are transparent.

\subsection{Generalised free energy and vector-field construction}

We begin with the generalised free energy
\begin{equation}\label{GFE}
\mathcal{F}(r_h,\tau) = M(r_h) - \frac{S(r_h)}{\tau},
\end{equation}
where $M$ and $S$ are the black-hole mass and entropy, and $\tau>0$
is an auxiliary parameter.  
When $\tau^{-1}$ is identified with the thermodynamic temperature, extrema of
$\mathcal{F}$ encode equilibrium conditions.  

The thermodynamic information is embedded in the two-component vector field
\begin{equation}\label{phi}
\boldsymbol{\phi}(r_h,\theta)
\;=\;
\begin{pmatrix}
\phi^{r_h} \\[4pt]
\phi^{\theta}
\end{pmatrix}
=
\begin{pmatrix}
\displaystyle\frac{\partial F}{\partial r_h} \\[8pt]
-\cot\theta\;\csc\theta
\end{pmatrix}.
\end{equation}
The first component vanishes at stationary points of $\mathcal{F}$,
and therefore captures the equilibrium condition.
The second component is chosen so that it vanishes only at
$\theta=\pi/2$, ensuring that zeros of $\boldsymbol{\phi}$
lie on the equatorial plane and correspond exactly to physical
thermodynamic equilibria.  
At such points one has $\tau^{-1}=T(r_h)$, i.e.\ the standard Hawking
temperature.  
Thus, the vector field \eqref{phi} singles out thermodynamic equilibrium
points as topological defects of the map $(r_h,\theta)\mapsto\boldsymbol{\phi}$.

\subsection{Normalised map and topological current}

To extract the topological information, we introduce the normalised
unit vector
\begin{equation}\label{eq:n-def}
n^a = \frac{\phi^a}{\phi},\qquad
\phi \equiv \sqrt{(\phi^{r_h})^2+(\phi^\theta)^2},
\qquad a=r_h,\theta .
\end{equation}
This defines a map from the parameter space into the unit circle $S^1$.
The associated topological current is
\begin{equation}\label{eq:top-current}
j^\mu
= \frac{1}{2\pi}\,\varepsilon^{\mu\nu\rho}\varepsilon^{ab}\,
\partial_\nu n_a\,\partial_\rho n_b,
\qquad
\mu,\nu,\rho=0,1,2,
\end{equation}
which is algebraically conserved: $\partial_\mu j^\mu=0$.  
Its time component,
\begin{equation}\label{eq:j0-def}
j^0
= \frac{1}{\pi}
\left(\partial_1 n_1\,\partial_2 n_2
      -\partial_2 n_1\,\partial_1 n_2\right),
\end{equation}
is the topological charge density measuring the local winding of the
map $(x^1,x^2)\!\mapsto\!(n_1,n_2)$.

In any region $D$ free of zeros of $\boldsymbol{\phi}$, the field $n_a$
is smooth and Eq.~\eqref{eq:j0-def} can be written as
\begin{equation}\label{eq:j0-div}
j^0 = \frac{1}{\pi}(\partial_1 Q - \partial_2 P),
\qquad
P=n_1\partial_1 n_2,\quad
Q=n_1\partial_2 n_2 .
\end{equation}
Using Green's theorem on a domain $D$ without defects,
\begin{equation}\label{eq:green}
\int_D j^0\,d^2x
= \frac{1}{\pi}\oint_{\partial D}(P\,dx^1 + Q\,dx^2)
= \frac{1}{\pi}\oint_{\partial D} n_1\,dn_2 = 0,
\end{equation}
since $n_a$ is single-valued there.
Thus, a non-zero topological charge can arise only when the integration
contour encloses a zero of $\boldsymbol{\phi}$.

\subsection{Winding number around isolated defects}

Let $C$ be a contour enclosing $N$ isolated zeros of $\boldsymbol{\phi}$
and let $c_i$ be small circles surrounding each zero.  
The total winding number is
\begin{equation}\label{eq:W-def}
W
= \frac{1}{\pi}\oint_C n_1\,dn_2
= \frac{1}{\pi}\sum_{i=1}^N \oint_{c_i} n_1\,dn_2 .
\end{equation}
To evaluate the integral around a single zero located at $(x_0,y_0)$,
we linearise the vector field as
\[
\phi^{r_h}(x,y)\approx f(x),\qquad
\phi^{\theta}(x,y)\approx g(y),
\]
with $f(x_0)=g(y_0)=0$ and $g'(y_0)$ rescaled to~1.
For a circular contour of radius $\varepsilon$,
$x=x_0+\varepsilon\cos t$, $y=y_0+\varepsilon\sin t$,
the leading behaviours are
\begin{eqnarray}\label{eq:local-expansions}
&&f(x(t)) = \varepsilon f'(x_0)\cos t + \mathcal{O}(\varepsilon^2),
\nonumber\\
&&g(y(t)) = \varepsilon\sin t + \mathcal{O}(\varepsilon^2).
\end{eqnarray}
Additionally, a direct computation gives the limiting integral
\begin{equation}\label{eq:circle-int-result}
\lim_{\varepsilon\to 0}
\oint_{c_\varepsilon} n_1\,dn_2
= \pi\,\frac{f'(x_0)}{|f'(x_0)|},
\qquad f'(x_0)\neq 0.
\end{equation}
Hence, each simple zero contributes $\pm\pi$ depending on the sign of
$f'(x_0)$, reflecting the rotation of the unit vector when circling the defect.

Returning to our variables, $\phi^{r_h}=0$ at a defect implies
$\partial_{r_h}\mathcal{F}=0$, and the derivative $f'(x_0)$ corresponds to
$\partial_{r_h}^2\mathcal{F}$ at that point.  
Therefore the total winding number is
\begin{equation}\label{eq:W-final}
W
= \sum_{i=1}^N
\operatorname{sgn}\!\left(
  \left.\frac{\partial^2\mathcal{F}}{\partial r_h^2}\right|_{r_h=r_i}
\right),
\end{equation}
where $r_i$ are the stationary points of $\mathcal{F}$.
Thus, the topology is completely characterised by the signs of the
second derivatives of the generalised free energy.

\subsection{Thermodynamic interpretation of the winding number}
\label{subsec:thermodynamic_winding_equivalence}

We now relate the winding number directly to thermodynamic stability
\cite{Silva:2025iip}.
From the equilibrium condition $\phi^{r_h}=0$ we obtain
\[
\tau = \frac{S'}{M'}, \qquad M'\neq 0,
\]
where primes denote $r_h$-derivatives evaluated at $r_h=r_i$.
The second derivative of $\mathcal{F}$ is
\begin{equation}\label{F2}
\left.\frac{\partial^2\mathcal{F}}{\partial r_h^2}\right|_{r_i}
= \frac{M''S' - M'S''}{S'}.
\end{equation}
For a monotonically increasing entropy ($S'>0$),
its sign is governed by $M''S' - M'S''$.

Now, the temperature and specific heat follow from
$T=\tfrac{dM}{dS}$ and $C=\tfrac{dM}{dT}$:
\begin{equation}\label{sC}
C = \frac{M' S'^2}{M''S' - M'S''},
\qquad M'>0.
\end{equation}
Thus,
\begin{equation}
\operatorname{sgn}(C)
= \operatorname{sgn}(M''S' - M'S'')
= \operatorname{sgn}\!\left(
\left.\frac{\partial^2\mathcal{F}}{\partial r_h^2}\right|_{r_i}
\right).
\end{equation}

We therefore obtain a direct correspondence:
\begin{eqnarray}
 \text{stable phase ($C>0$)} \;&\Longleftrightarrow&\; w_i=+1,
\nonumber\\ 
\text{unstable phase ($C<0$)} \;&\Longleftrightarrow&\; w_i=-1.
\end{eqnarray}
 Hence, the thermodynamic stability of each branch is encoded in the
topological index of the corresponding defect, establishing a clean link
between thermodynamic and topological descriptions of black-hole
equilibria.

\section{Topological classification and photon-sphere analysis: general 
formalism}
\label{Sec:TopologicalClassificationGeneral}

In this section  we develop the general framework used to classify the 
thermodynamic topology of entropy-deformed black holes and to study the 
associated photon-sphere structure. Our starting point is the entropy-geometry 
correspondence introduced in \cite{Anand:2025rjg}, which shows that any 
modification of the entropy induces a corresponding backreaction on the metric. 
This framework allows us to treat entropy as the fundamental input and derive 
both the geometric and thermodynamic behaviour in a unified manner. We first 
present the thermodynamic and topological classification in full generality, 
and then analyse the photon-sphere properties implied by the modified metric.
This reformulation makes explicit that all thermodynamic and geometric 
observables follow directly from the chosen entropy model, allowing us to 
classify modified-black-hole physics in a fully entropy-driven and 
model-independent manner.

\subsection{Thermodynamic and topological classification}

We begin with the general static and spherically symmetric line element
\begin{equation}
    ds^2 = -f(r)\,dt^2 
           + \frac{dr^2}{f(r)} 
           + r^2 d\Omega^2 ,
\end{equation}
and parametrize the lapse function as
\begin{equation}
   f(r) = 1 - M g(r) .
\end{equation}
The backreaction induced by an arbitrary entropy function $S(r)$ can be encoded 
in the general metric constructed in \cite{Anand:2025rjg}, namely
\begin{equation}\label{metric_gen}
    ds^2 = -\left(1 - \frac{4\pi M}{S'(r)}\right) dt^2 
           + \frac{dr^2}{\left(1 - \frac{4\pi M}{S'(r)}\right)} 
           + r^2 d\Omega^2 ,
\end{equation}
so that the metric function is expressed as
\begin{equation}
    f(r) = 1 - \frac{4\pi M}{S'(r)} .
\end{equation}
The horizon radius $r_h$ satisfies $f(r_h)=0$, and the corresponding Hawking 
temperature is
\begin{equation}
    T_H = \frac{S''(r_h)}{4\pi\,S'(r_h)} .
\end{equation}
Furthermore, the generalized off-shell free-energy density is given by
\begin{equation}\label{Off_shell_free_energy_def}
    \mathcal{F}(r_h,\tau) 
    = \frac{S'(r_h)}{4\pi} - \frac{S(r_h)}{\tau} ,
\end{equation}
from which the components of the $\phi$-mapping vector field follow as
\begin{equation}
    \phi^{r_h} 
      = \frac{S''(r_h)}{4\pi} - \frac{S'(r_h)}{\tau},
    \qquad
    \phi^{\theta} = -\cot\theta\,\csc\theta .
\end{equation}
The condition $\phi^{r_h}=0$ determines the critical radius $r_i$ through
\begin{equation}\label{Tau_gen}
    \frac{S''(r_i)}{S'(r_i)} = \frac{4\pi}{\tau} .
\end{equation}

To classify the nature of the thermodynamic critical point, we evaluate the 
second derivative of the free energy at $r_i$:
\begin{equation}\label{d2F_general}
   \left.\frac{\partial^2\mathcal{F}}{\partial r_h^2}\right|_{r_h=r_i}
   = -\frac{S''(r_i)^2 - S^{(3)}(r_i)\,S'(r_i)}{4\pi\,S'(r_i)} .
\end{equation}
Its sign determines the winding number associated with the topological defect: 
positive for stable sectors and negative for unstable ones. The dependence on 
$S$, $S'$ and $S''$ shows that the topology is fully controlled by the entropy 
expression rather than by the matter content.

The equivalence with the residue method follows directly. Using the horizon 
condition applied to \eqref{metric_gen}, the mass can be written as
\begin{equation}
    M = \frac{S'(r_h)}{4\pi} ,
\end{equation}
and differentiation yields
\begin{equation}\label{GenResidu}
    M'' S' - M' S'' = 
    -\frac{S''(r_i)^2 - S^{(3)}(r_i)\,S'(r_i)}{4\pi} .
\end{equation}
Thus, the sign of the residue in the mass-temperature plane matches the sign of 
the second derivative of the free energy, confirming that both approaches 
encode 
the same topological information.

\subsection{Photon analysis and constraints}

As shown in \cite{Qiao:2022jlu,Qiao:2022hfv}, the photon sphere of a static and 
spherically symmetric spacetime admits a clear geometric characterization once 
the optical metric is introduced via $ds^{2}=0$ and $dt^{2} = g^{\rm OP}_{ij} 
dx^{i} dx^{j}$. Two quantities play a central role here: the geodesic curvature 
of 
circular null orbits and the Gaussian curvature of the optical 2-geometry.

The photon sphere radius $r_{ph}$ is determined by the vanishing of the 
geodesic 
curvature,
\begin{equation}\label{geodesic_curvature}
	\kappa_{g} = 
 \left[\frac{f(r)}{r} - \frac{1}{2} f'(r)\right]_{r=r_{ph}} = 0 .
\end{equation}
Moreover, the Gaussian curvature of the optical metric is
\begin{equation}\label{Gauss_curvature}
	\mathcal{K} 
 = \frac{1}{2} f(r)\, f''(r)
   - \left[\frac{1}{2} f'(r)\right]^{2},
\end{equation}
with the standard stability conditions
\begin{align}
	\mathcal{K} < 0 &: \text{unstable photon orbit}, \nonumber\\
	\mathcal{K} > 0 &: \text{stable photon orbit}. \nonumber
\end{align}

Now, using the entropy-deformed lapse function \eqref{metric_gen}, the 
geodesic 
curvature becomes
\begin{equation}
    \kappa_g
    = -\frac{4\pi M}{r\,S'(r)}
      - \frac{2\pi M\,S''(r)}{S'(r)^2}
      + \frac{1}{r} .
\end{equation}
In order to highlight the effect of modified entropy, we write
\begin{equation}\label{ent}
    S = S_{\rm BH} + \mathcal{S}(A),
    \qquad S_{\rm BH} = \pi r^{2},
\end{equation}
where $\mathcal{S}(A)$ encodes deviations from the Bekenstein-Hawking law. 
Hence, from the above we finally obtain  
\begin{equation}
    \kappa_g
    = -\frac{2M}{r^{2}\left[1+4\frac{\partial\mathcal{S}}{\partial A}\right]}
      - \frac{M\left[\left(1+4\frac{\partial\mathcal{S}}{\partial A}\right)
      + 4r\,\frac{\partial}{\partial r}
          \left(\frac{\partial\mathcal{S}}{\partial A}\right)\right]}
      {r^{2}\left[1+4\frac{\partial\mathcal{S}}{\partial A}\right]^{2}}
      + \frac{1}{r} .
\end{equation}
This expression shows explicitly how entropy modifications shift the photon 
sphere radius, and in particular one can see that only the derivative of the 
entropy   enters, which implies that different generalized entropies lead to 
distinct optical signatures.
Finally, setting $\partial\mathcal{S}/\partial A =0$ 
yields the 
Schwarzschild limit 
\begin{equation}
    \kappa_g = \frac{1}{r} - \frac{3M}{r^{2}},
\end{equation}
which vanishes at $r_{ph} = 3M$.

Using Eq.~\eqref{Gauss_curvature}, the Gaussian curvature becomes
\begin{eqnarray}
	\mathcal{K} & = & -\frac{8 \pi ^2 M^2 S^{(3)}(r)}{S'(r)^3}+\frac{12 \pi ^2 
M^2 S''(r)^2}{S'(r)^4} \nonumber \\
    &+& \frac{2 \pi  M S^{(3)}(r)}{S'(r)^2}-\frac{4 \pi  M S''(r)^2}{S'(r)^3},
\end{eqnarray}
and therefore substituting the decomposition \eqref{ent}, we obtain
\begin{equation}
\mathcal{K}
= \frac{8M^{2}\mathcal{X}(r)}{r^{4}\left(1 + 
4\frac{\partial\mathcal{S}}{\partial A}\right)^{4}}
 + \frac{M\,\mathcal{Y}(r)}{r^{2}\left(1 + 4\frac{\partial\mathcal{S}}{\partial 
A}\right)^{3}}
 - \frac{2M}{r^{3}\left(1 + 4\frac{\partial\mathcal{S}}{\partial A}\right)},
\end{equation}

where 
\begin{eqnarray}\notag
&&\mathcal{X}(r)= -4 r^2 \left(\frac{1}{4}+\frac{\partial \mathcal{S}}{\partial 
A} \right) \frac{\partial^2}{\partial r^2} \left(\frac{\partial 
\mathcal{S}}{\partial A}\right) +6 \left(\frac{1}{4}+ \frac{\partial 
\mathcal{S}}{\partial A}\right)^2\\
&+&6 r^2  \left[\frac{\partial}{\partial r} \left(\frac{\partial 
\mathcal{S}}{\partial A}\right)\right]^2 
+ r \left(1+4 \frac{\partial \mathcal{S}}{\partial A}\right) 
\left[\frac{\partial}{\partial r} \left(\frac{\partial \mathcal{S}}{\partial 
A}\right)\right],
\end{eqnarray}
and 
\begin{eqnarray}\notag
\mathcal{Y}(r)&=& 4 r \left(1+4\frac{\partial \mathcal{S}}{\partial A} \right) 
\frac{\partial^2}{\partial r^2} \left(\frac{\partial \mathcal{S}}{\partial 
A}\right) - 8 r  \left[\frac{\partial}{\partial r} \left(\frac{\partial 
\mathcal{S}}{\partial A}\right)\right]^2 \\\notag
&&- 2\left(1+4 \frac{\partial \mathcal{S}}{\partial A}\right) 
\left[\frac{\partial}{\partial r} \left(\frac{\partial \mathcal{S}}{\partial 
A}\right)\right].
\end{eqnarray}

As we can see, in the Bekenstein-Hawking case ($\mathcal{S}(A)=0$), the 
Gaussian optical 
curvature
of the equatorial optical metric becomes
\begin{equation}
    \mathcal{K} = -\frac{2M}{r^{3}} + \frac{3M^{2}}{r^{4}} .
\end{equation}
Evaluating this at the Schwarzschild photon-sphere radius $r_{ph}=3M$, we obtain
\begin{eqnarray}
    \mathcal{K}(r_{ph}) 
    &=& - \frac{1}{27 M^2}
     = -\frac{1}{r_{sh}^2} ,
\end{eqnarray}
where we have used the exact relation $r_{sh}=3\sqrt{3}\,M$ for the 
Schwarzschild
shadow.  Thus, for the standard entropy law, the shadow radius satisfies
\begin{equation}
    r_{sh} = \frac{1}{\sqrt{|\mathcal{K}(r_{ph})|}} .
\end{equation}
The above  relation is exact in the undeformed (Schwarzschild) case. For 
modified
entropies, and therefore modified metrics, the same expression continues to
provide an accurate estimate for $r_{sh}$, since the optical geometry remains a
small perturbation of the Schwarzschild one in the neighborhood of the photon
sphere.  We therefore use this curvature-based expression as a controlled
approximation when evaluating the shadow radius in the generalized entropy
models considered below.
 Finally, in the general case, the above relation is only approximate, since
\begin{equation}
	r_{sh} = \frac{r_{ph}}{\sqrt{f(r_{ph})}}
    \;\sim\; \frac{1}{\sqrt{|\mathcal{K}|}} + \mathcal{O}(M).
\end{equation}

Lastly, we employ observational constraints to assess the possible signatures 
of modified entropies. Using EHT observations of Sgr A* and setting $M=1$, the 
shadow-radius bounds at the $2\sigma$ level are \cite{Vagnozzi:2022moj}
\begin{equation}\label{const}
    4.21 \lesssim r_{\rm sh} \lesssim 5.56 .
\end{equation}
These constraints will be used to bound the deformation parameters of the 
specific entropy expressions analysed in the next section.

\section{Topological classification and photon-sphere analysis in specific 
modified entropy cases}\label{SecIV}

In the previous sections we established the general topological framework for 
entropy-deformed black holes and demonstrated how modified thermodynamic 
potentials give rise to distinct topological charges, which in turn classify 
the 
equilibrium structure of the solutions. We also showed how these thermodynamic 
features leave imprints on the photon sphere and the associated optical 
geometry, allowing observational quantities such as the shadow radius to encode 
signatures of entropy corrections. In this section, we apply the general 
thermodynamic-topology and photon-sphere
formalism to four well-motivated entropy deformations that appear in 
quantum gravity, statistical mechanics, and gravitational thermodynamics.
These four entropy deformations represent distinct classes of quantum or 
statistical corrections, namely fractal (Barrow), non-extensive (R\'enyi), 
non-Gaussian relativistic (Kaniadakis), and quantum-loop/string-motivated 
(logarithmic) ones, allowing for a unified comparison within the same 
topological 
and optical framework.

\subsection{Barrow entropy}

Barrow proposed a fractal deformation of the Bekenstein-Hawking entropy, 
motivated by the possibility that quantum-gravitational effects induce a 
microscopic fractalisation of the horizon surface \cite{Barrow:2020tzx}. The 
modified entropy is characterised by a single parameter \(\Delta\), which 
quantifies the degree of fractality, and is given by
\begin{equation}\label{Barrow_ent_def}
S_{\mathrm{B}} = \left(\pi r_h^{2}\right)^{1+\Delta/2},
\end{equation}
with the theoretical range
\[
0 \leq \Delta \leq 1.
\]
For \(\Delta=0\), the standard area law is recovered, while increasing 
\(\Delta\) encodes progressively stronger deviations from smooth horizon 
geometry.

\subsubsection{Thermodynamic and topological classification}

Using the generalised off-shell free energy \eqref{Off_shell_free_energy_def}, 
the Barrow free-energy functional becomes
\begin{equation}\label{GenF_Barrow}
    \mathcal{F}_B
    = \frac{1}{4}\,\pi^{\Delta/2}\,(\Delta+2)\,r_h^{\Delta+1}
      - \frac{\pi^{1+\Delta/2}\,r_h^{\Delta+2}}{\tau}.
\end{equation}
Applying the equilibrium condition $\phi^{r_h}=0$ from Eq.~\eqref{Tau_gen} 
yields the auxiliary parameter
\begin{equation}\label{Tau_barrow}
    \tau_B = \frac{4\pi\, r_i}{\Delta+1},
\end{equation}
and therefore the critical radius satisfies
\begin{equation}
    r_i = \frac{(\Delta+1)\,\tau_B}{4\pi}.
\end{equation}

To determine the local topological index, we evaluate the second derivative of 
the free energy at $r_i$:
\begin{equation}\label{d2F_Barrow}
    \left.\frac{\partial^{2}\mathcal{F}_B}{\partial r_h^{2}}\right|_{r_h=r_i}
    = 
-\frac{4^{-\Delta}\,\pi^{1-\Delta/2}\,(\Delta+2)\,[(\Delta+1)\tau]^{\Delta}}{
\tau}.
\end{equation}
Since \(\Delta>0\) and \(r_i>0\), this quantity is strictly negative. 
Consequently,
\begin{equation}
    w
    = \operatorname{sgn}\!\left[
        \left.\frac{\partial^{2}\mathcal{F}_B}{\partial 
r_h^{2}}\right|_{r_h=r_i}
      \right]
    = -1.
\end{equation}
This shows that Barrow-modified black holes correspond to a locally unstable 
thermodynamic branch at the equilibrium radius \(r_i\). Finally, using 
 \eqref{eq:W-final}, we find that  the global topological charge for the Barrow 
entropy 
  is therefore \(W=-1\).

\subsubsection{Photon-sphere analysis and constraints from Sgr A*}

In order to analyse the optical properties of Barrow entropy, we adopt the 
Barrow-corrected metric function obtained in \cite{Anand:2025rjg},
\begin{equation}
    f_{B}(r)
    = 1 - \frac{4\pi^{-\Delta/2}\,M}{(\Delta+2)\,r^{\Delta+1}},
\end{equation}
with \(M\) the  Arnowitt-Deser-Misner (ADM) mass. For 
$0<\Delta\ll 1$, this supports a perturbative 
expansion.

Using  relation \eqref{geodesic_curvature}, the geodesic curvature becomes
\begin{equation}
    \kappa_g^{B}(r)
    = -\frac{6\pi^{-\Delta/2} M}{(\Delta+2)\,r^{\Delta+2}}
      - \frac{2\Delta\,\pi^{-\Delta/2}M}{(\Delta+2)\,r^{\Delta+2}}
      + \frac{1}{r}.
\end{equation}
Solving $\kappa_g^{B}(r_{ph}^{B})=0$ gives the photon-sphere location. 
Then, expanding 
to linear order in \(\Delta\) yields
\begin{equation}
    r_{ph}^{B}
    = 3M
    - \frac{M}{2}\left[1 + \ln\!\big(729\pi^{3}M^{6}\big)\right]\Delta
    + \mathcal{O}(\Delta^{2}).
\end{equation}
Hence, the leading modification introduces a logarithmic dependence on the 
mass 
scale, while the Schwarzschild result \(r_{ph}=3M\) is recovered in the limit 
\(\Delta\to 
0\).

The Gaussian optical curvature from relation \eqref{Gauss_curvature}, evaluated 
at 
the perturbed photon radius, becomes
\begin{equation}
    \mathcal{K}^{B}(r_{ph}^{B})
    = -\frac{1}{27M^{2}}
      - \frac{2+\ln(9\pi M^{2})}{27M^{2}}\,\Delta
      + \mathcal{O}(\Delta^{2}),
\end{equation}
which remains negative, confirming the instability of the null circular orbit.
Moreover, the corresponding shadow radius is
\begin{equation}
    r_{sh}^{B}
    = 3\sqrt{3}\,M
      - \frac{3\sqrt{3}\,M}{2}\left[1+\ln(9\pi M^{2})\right]\Delta
      + \mathcal{O}(\Delta^{2}).
\end{equation}
Thus, Barrow entropy predicts an $\mathcal{O}(\Delta)$ reduction in the shadow 
size compared to the Schwarzschild value.

\begin{figure}[ht]
\centering
\includegraphics[scale=0.6]{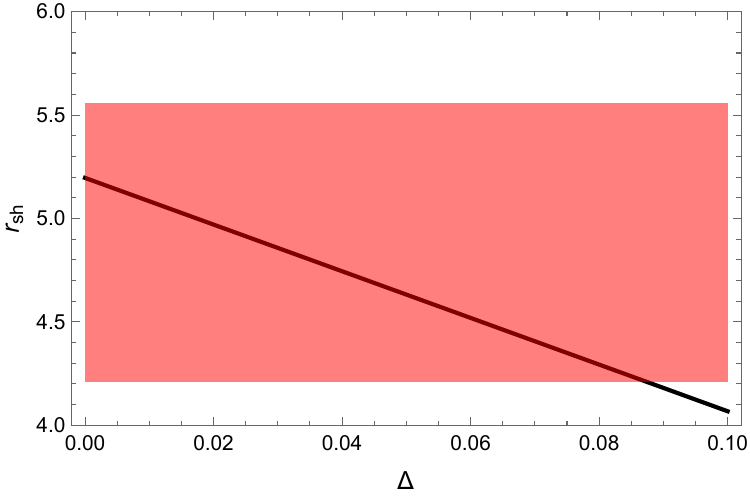}
\caption{{\it{Shadow radius for the Barrow-entropy-corrected black hole. The 
red region is consistent with the EHT horizon-scale image of Sgr A* at $2 
\sigma 
$. }}}
\label{shadow1}
\end{figure}

Lastly, using the EHT $2\sigma$ bounds in \eqref{const} for Sgr A*, and 
setting 
$M=1$, the Barrow parameter is constrained to
\begin{equation}
    \Delta \lesssim 0.08744.
\end{equation}
These results are shown in Fig. \ref{shadow1}.
The obtained constraint is consistent with bounds obtained from Big Bang 
Nucleosynthesis 
\cite{Barrow:2020kug,Sheykhi:2024fya}, stellar-orbit measurements 
\cite{Jusufi:2021fek}, and cosmological datasets 
\cite{Leon:2021wyx,Asghari:2021bqa,Luciano:2025hjn}.

\subsection{R\'enyi entropy}

R\'enyi entropy is a one-parameter generalisation of the Shannon 
(Boltzmann-Gibbs) entropy and is extensively used in information theory, 
statistical mechanics, and gravitational thermodynamics \cite{Renyi:1961,aa,bb} 
. In particular, for a discrete 
probability distribution \(\{p_i\}_{i=1}^W\) with \(p_i\ge 0\) and \(\sum_i p_i 
= 1\), the R\'enyi entropy of order \(\lambda\) is defined as
\begin{equation}\label{eq:Renyi-def}
S_{\mathrm{R},\lambda}
= \frac{1}{1-\lambda}\,
  \ln\!\left(\sum_{i=1}^W p_i^{\,\lambda}\right),
\qquad \lambda \in \mathbb{R},\ \lambda \neq 1.
\end{equation}
Thus, in the limit \(\lambda\to 1\), one recovers the standard Shannon 
entropy. For 
black holes, the R\'enyi entropy takes the form
\begin{equation}\label{Renyi_Entropy}
    S_R = \frac{\log\!\left(1+\lambda\,\pi\,r_h^2\right)}{\lambda},
\end{equation}
where \(\lambda\) is the R\'enyi deformation parameter. Finally, for 
thermodynamic 
consistency, the commonly adopted interval is
\[
0 < \lambda \lesssim 1,
\]
ensuring positivity, concavity, and a smooth Bekenstein-Hawking limit as 
\(\lambda\to 0\).

\subsubsection{Thermodynamic and topological classification}

\begin{figure}[ht]
\centering
\includegraphics[scale=0.45]{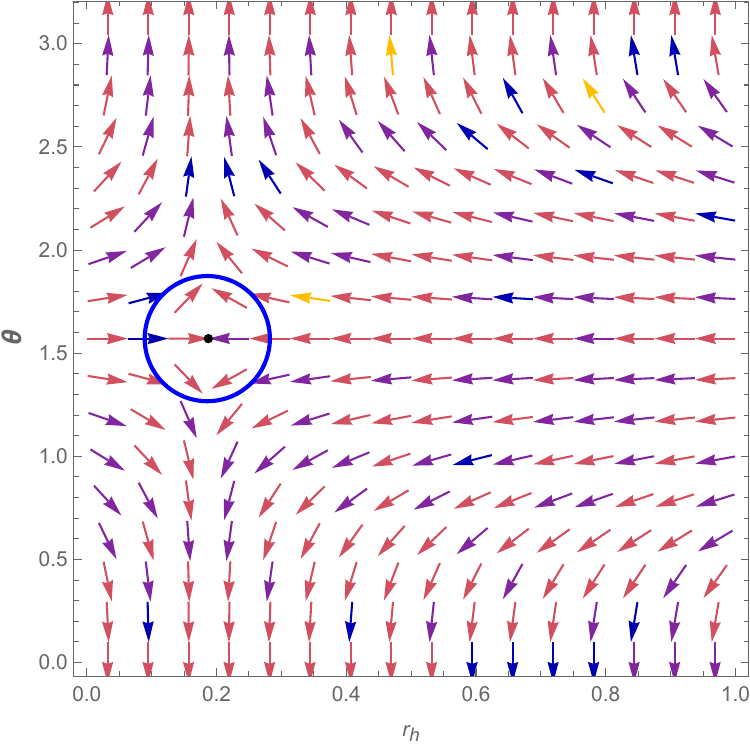}
\caption{{\it{ The $r_h-\theta$ diagram in the case of  R\'enyi-modified black 
hole entropy. As we observe, there is only one winding number, namely 
$w_1=-1$.}}}
\label{fig:Reny_r_theta}
\end{figure}

Using  \eqref{Off_shell_free_energy_def} together with 
\eqref{Renyi_Entropy}, 
the generalised free energy becomes
\begin{equation}
{\cal F}_R(r_h)
= \frac{r_h}{2(\pi \lambda r_h^2 + 1)}
  - \frac{\ln\!\big(1+\pi \lambda r_h^2\big)}{\lambda \tau}.
\end{equation}
The first and second derivatives with respect to \(r_h\) are
\begin{align}
\frac{\partial {\cal F}}{\partial r_h}
 &= \frac{\tau - \pi r_h\left[\lambda r_h(4\pi r_h+\tau)+4\right]}
         {2\tau\left(\pi \lambda r_h^2 + 1\right)^2},  \\[4pt]
\label{d2F_Renyi}
\frac{\partial^{2}{\cal F}}{\partial r_h^{2}}
 &= \frac{\pi\left[\lambda r_h\left(\pi\lambda r_h^2 (2\pi r_h+\tau) - 
3\tau\right)-2\right]}
         {\tau\left(\pi \lambda r_h^2 + 1\right)^3}.
\end{align}
Imposing the equilibrium condition \(\phi^{r_h}=0\) via \eqref{Tau_gen} 
yields
\begin{equation}\label{Tau_Renyi}
    \tau_R
    = 4\pi r_i\left(\frac{2}{1 - \pi\lambda r_i^2} - 1\right).
\end{equation}
In order to analyse the roots of this equation, it is convenient to rewrite it 
as the 
cubic polynomial
\begin{equation}
   f_R(r_i)
   = r_i^3 + \frac{\tau}{4\pi} r_i^2
     + \frac{r_i}{\pi\lambda}
     - \frac{\tau}{4\pi^{2}\lambda}.
\end{equation}
Since \(\partial_{r_i} f_R(r_i)\) is strictly increasing and
\[
\lim_{r_i\to 0} f_R(r_i) = -\frac{\tau}{4\pi^{2}\lambda},
\qquad
\lim_{r_i\to\infty} f_R(r_i) = +\infty,
\]
the intermediate value theorem guarantees exactly one positive root \(r_i\). 
Moreover, in 
the perturbative regime, the critical radius expands as
\begin{equation}
    r_i
    = \frac{\tau_R}{4\pi}
      + \frac{5\sqrt{\lambda}\,\tau^{2}}{96\sqrt{3}\,\pi^{3/2}}
      - \frac{11\lambda\,\tau^{3}}{576\pi^{2}}
      + \mathcal{O}(\lambda^{2}).
\end{equation} 
Substituting  \eqref{Tau_Renyi} into \eqref{d2F_Renyi}, the curvature of the 
free energy evaluated at \(r_i\) is
\begin{equation}
\left.\frac{\partial^{2}{\cal F}_R}{\partial r_h^{2}}\right|_{r_h=r_i}
= -\frac{2\pi}{\tau}
  - \frac{3\lambda\,\tau}{8}
  + \mathcal{O}(\lambda^{2}).
\end{equation}
This quantity is negative for all admissible values of \(\lambda\), yielding 
the 
winding number
\[
w = -1.
\]
Thus, the R\'enyi-modified black hole belongs to the same topological class as 
the 
Barrow case, and the total topological charge is
\begin{equation}
    W = \sum_{i=1}^{N} w_i = -1.
\end{equation}
In order to  evaluate the winding number, we can also analyse the structure of 
the vector field \(\phi\) in the \((r_h,\theta)\) plane, and the corresponding   
 $r_h-\theta$ diagram  is   illustrated in Fig.~\ref{fig:Reny_r_theta}. From 
this diagram we also conclude that  there is only one winding number, namely 
$w_1=-1$.

\subsubsection{Photon-sphere analysis and constraints from Sgr A*}

We proceed by studying the possible  observational signatures of the 
aforementioned analysis. We adopt the R\'enyi-corrected metric 
function introduced in \cite{Anand:2025rjg},
\begin{equation}
    f_{R}(r)
    = 1 - \frac{2M\left(1+\pi\lambda r^{2}\right)}{r},
\end{equation}
where \(M\) is the ADM mass. Thus, for \(0<\lambda\ll 1\)  we can work 
perturbatively. 
Substituting \(f_R(r)\) into the geodesic curvature formula 
\eqref{geodesic_curvature} yields
\begin{equation}
    \kappa_g^{R}(r)
    = -\pi\lambda M
      - \frac{3M}{r^{2}}
      + \frac{1}{r}.
\end{equation}
Setting \(\kappa_g^{R}(r_{ph}^{R}) = 0\) gives the photon-sphere radius
\begin{equation}
    r_{ph}^{R}
    = 3M
      + 9\pi\lambda M^{3}
      + \mathcal{O}(\lambda^{2}).
\end{equation}
The Schwarzschild value is recovered at \(\lambda=0\), while R\'enyi 
corrections 
shift the photon sphere outward.

The Gaussian optical curvature \eqref{Gauss_curvature}, evaluated at the 
perturbed photon radius, becomes
\begin{equation}
    \mathcal{K}^{R}(r_{ph}^{R})
    = -\frac{1}{27M^{2}}
      + \frac{8\pi\lambda}{9}
      + \mathcal{O}(\lambda^{2}),
\end{equation}
which remains negative for all relevant values of \(\lambda\), confirming that 
the circular null orbit continues to be unstable.
Additionally, the corresponding shadow radius is
\begin{equation}
    r_{sh}^{R}
    = 3\sqrt{3}\,M
      + 27\pi\sqrt{3}\,\lambda M^{3}
      + \mathcal{O}(\lambda^{2}),
\end{equation}
indicating an \(\mathcal{O}(\lambda)\) enlargement of the shadow relative to 
Schwarzschild.

\begin{figure}
\centering
\includegraphics[scale=0.6]{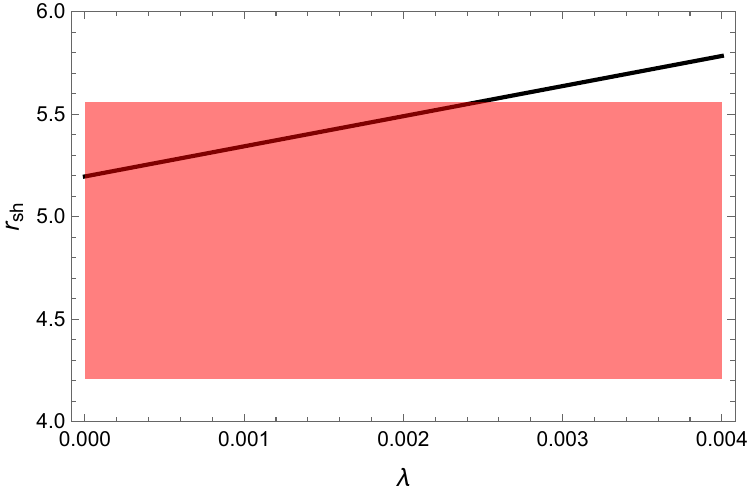}
\caption{{\it{Shadow radius for the R\'enyi-entropy-corrected black hole. The 
red 
region is consistent with the EHT horizon-scale image of Sgr A at $2 \sigma 
$.}}}
\label{shadow2}
\end{figure}

Now, using the $2\sigma$ EHT constraint \eqref{const} for Sgr A* with \(M=1\), 
we find that  the R\'enyi parameter must satisfy
\begin{equation}
    \lambda \lesssim 0.00248.
\end{equation}
These results are presented in Fig. \ref{shadow2}.
This constraint is compatible with bounds from primordial Big-Bang 
Nucleosynthesis and baryogenesis 
\cite{sheykhi2025constraintsrenyientropyprimordial}.

\subsection{Logarithmically corrected entropy}

Logarithmic corrections to the Bekenstein-Hawking area law arise universally 
in 
a wide range of quantum-gravity frameworks. These include Loop Quantum Gravity 
(via counting of spin-network punctures), string-theoretic one-loop 
corrections, 
heat-kernel expansions in Euclidean quantum gravity, entanglement entropy of 
quantum fields across the horizon, and conformal field theory approaches 
\cite{Kaul:2000kf,Das:2001ic, Sen:2012dw}. In 
all these cases, quantum fluctuations around the classical saddle introduce 
subleading terms proportional to \(\log(\pi r_h^{2})\).

A generic and widely used form of the corrected entropy is
\begin{equation}\label{eq:log_entropy}
S_{\rm log}(r_h)
 = \pi r_h^{2}
 + \lambda\,\log\!\left(\pi r_h^{2}\right),
\end{equation}
where the parameter \(\lambda\) encodes the magnitude and sign of the quantum 
correction. Its sign is theory-dependent: Loop Quantum Gravity typically 
predicts \(\lambda<0\), while string theory and CFT-based methods may produce 
either sign. The sign of \(\lambda\) affects the small-radius thermodynamics 
and 
the existence of stable or unstable branches.

\subsubsection{Thermodynamic and topological classification}

Substituting \eqref{eq:log_entropy} into the generalised free energy 
\eqref{Off_shell_free_energy_def} gives
\begin{equation}\label{d2f_log}
    {\cal F}_{\rm log}(r_h)
    = \frac{\lambda + \pi r_h^{2}}{2\pi r_h}
      - \frac{\pi r_h^{2} + \lambda \log\!\left(\pi r_h^{2}\right)}{\tau}.
\end{equation}
The thermodynamic equilibrium is determined by the stationary condition
\[
\left.\frac{\partial {\cal F}_{\rm log}}{\partial r_h}\right|_{r_h=r_i}=0,
\]
which yields the relation
\begin{equation}
    r_i^{2}
    = \frac{\lambda}{\pi}
      + \frac{4 r_i\left(\lambda + \pi r_i^{2}\right)}{\tau}.
\end{equation}
Solving for \(\tau\) gives the auxiliary parameter
\begin{equation}\label{Tau_log}
    \tau_{\rm log}
    = \frac{4\pi r_i\left(\lambda + \pi r_i^{2}\right)}
           {\pi r_i^{2} - \lambda},
\end{equation}
and rewriting \eqref{Tau_log} in cubic form, namely
\begin{equation}
    f_{\rm log}(r_i)
    = r_i^{3}
      - \frac{\tau_{\rm log}}{4\pi}\,r_i^{2}
      + \frac{\lambda}{\pi}\,r_i
      + \frac{\lambda \tau_{\rm log}}{4\pi^{2}},
\end{equation}
makes the structure of solutions more transparent. Since the leading term is 
positive, we have
\[
f_{\rm log}(0) = \frac{\lambda \tau_{\rm log}}{4\pi^{2}}, \qquad
\lim_{r_i\to\infty} f_{\rm log}(r_i) = +\infty.
\]
Thus, the cubic form begins positive at \(r_i=0\) and remains positive for 
sufficiently large \(r_i\). By continuity, this implies that the equation may 
admit either zero or two positive real roots, but never exactly one. 
These different possibilities correspond to the appearance of distinct 
thermodynamic branches.

\begin{figure}[ht]
\centering
\includegraphics[scale=0.45]{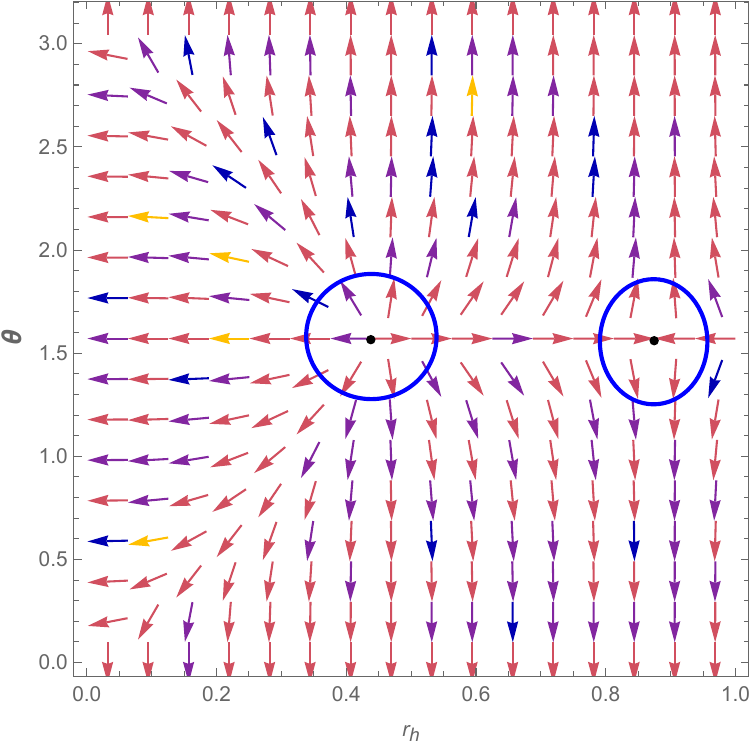}
\caption{{\it{The $r_h-\theta$ diagram in the case of logarithmically corrected 
black-hole entropy. As we observe, there exist two different values of 
critical 
radius at fixed $\tau$, therefore at different critical radii we have different 
winding numbers, namely $w_i=+1,-1$ respectively.}}}
\label{Fig:Log_R_theta}
\end{figure}

In Fig. \ref{Fig:Log_R_theta} we present  the  corresponding phase portrait. 
As we observe, 
the logarithmic correction permits both stable and unstable equilibrium points. 
The system therefore admits winding numbers of opposite sign at different radii 
\(r_i\), and as a consequence, the total topological charge becomes
\begin{equation}
    W = w_1 + w_2 = 0,
\end{equation}
indicating a cancellation between the two topological sectors. This behaviour 
is 
qualitatively different from the Barrow and R\'enyi cases, where the corrected 
theories yielded a net topological charge of $-1$.

\begin{figure}[ht]
\centering
\includegraphics[scale=0.6]{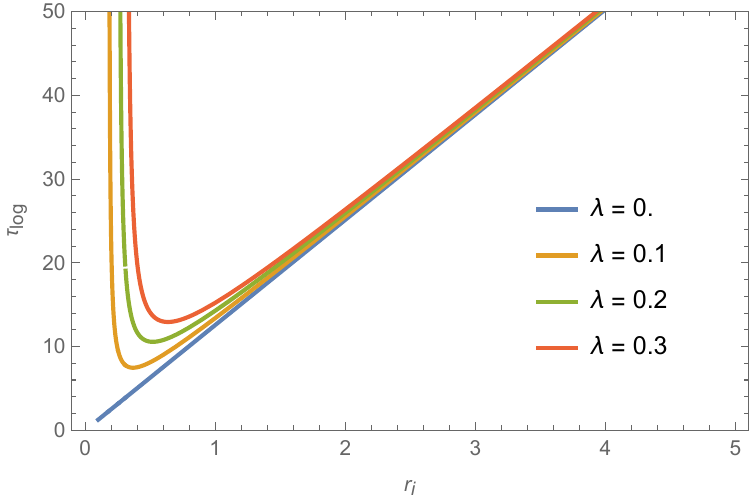}
\caption{{\it{ The auxiliary temperature   parameter $\tau_{\rm log}$ versus 
the critical radius $r_i$,
in the case of logarithmically corrected 
black-hole entropy, for 
different 
$\lambda$ values. When $\lambda=0$, the behaviour reduces to the Schwarzschild 
case with a single critical radius. For nonzero $\lambda$, the curve splits 
into 
two intersections at the same $\tau_{\rm log}$, revealing the emergence of two 
distinct critical radii and associated topological branches.}}}
\label{Fig:Tau_log}
\end{figure}

In  Fig. \ref{Fig:Tau_log} we depict the auxiliary temperature   parameter 
$\tau_{\rm log}$ versus the critical radius $r_i$ for 
different values of 
$\lambda$. As we can see, it is clear that for $\lambda=0$  the spacetime is 
the limiting case of Schwarzschild solution. However,  beyond that we have two 
limiting 
values, corresponding to the winding numbers obtained in 
Fig.~\ref{Fig:Log_R_theta}.

\subsubsection{Photon-sphere analysis and constraints from Sgr A*}

For the optical properties, we employ the log-corrected metric function 
proposed 
in \cite{Anand:2025rjg},
\begin{equation}
    f_{\rm log}(r)
    = 1 - \frac{2M\pi r}{\lambda + \pi r^{2}}.
\end{equation}
Substituting this expression into  \eqref{geodesic_curvature}, the geodesic 
curvature becomes
\begin{equation}
    \kappa_{g}^{\rm log}(r)
    = -\frac{3\pi^{2} M r^{2}}{\big(\lambda + \pi r^{2}\big)^{2}}
      - \frac{\pi \lambda M}{\big(\lambda + \pi r^{2}\big)^{2}}
      + \frac{1}{r}.
\end{equation}
Imposing the photon-sphere condition \(\kappa_{g}^{\rm log}(r_{ph}^{\rm 
log})=0\) and expanding to linear order in \(\lambda\) yields
\begin{equation}
    r_{ph}^{\rm log}
    = 3M
      - \frac{5\lambda}{9\pi M}
      + \mathcal{O}(\lambda^{2}).
\end{equation}
Thus, the Schwarzschild value \(r_{ph}=3M\) is recovered for \(\lambda=0\), 
while logarithmic corrections induce a shift governed by the ratio 
\(\lambda/M\).
Evaluating the Gaussian curvature via \eqref{Gauss_curvature} at the 
perturbed photon radius gives
\begin{equation}
    \mathcal{K}^{\rm log}(r_{ph}^{\rm log})
    = -\frac{1}{27M^{2}}
      + \frac{4\lambda}{729\pi M^{4}}
      + \mathcal{O}(\lambda^{2}),
\end{equation}
which remains negative, confirming that the circular photon orbit continues to 
be unstable.
Additionally, the resulting shadow radius is
\begin{equation}
    r_{sh}^{\rm log}
    = 3\sqrt{3}\,M
      - \frac{\lambda}{\sqrt{3}\,\pi M}
      + \mathcal{O}(\lambda^{2}),
\end{equation}
showing that logarithmic entropy corrections decrease the shadow size at linear 
order in \(\lambda\).

\begin{figure}
\centering
\includegraphics[scale=0.6]{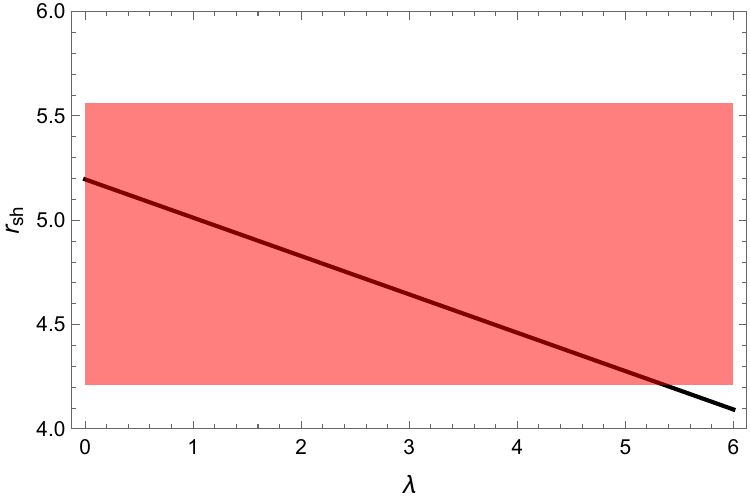}
\caption{{\it{Shadow radius for the logarithmically-corrected   
  black hole. The red region 
is consistent with the EHT horizon-scale image of Sgr A at $2 \sigma $.}}}
\label{shadow3}
\end{figure}

Finally, using the EHT \(2\sigma\) constraint \eqref{const} for Sgr~A$^{\ast}$ 
with 
\(M=1\), the logarithmic parameter is bounded by
\begin{equation}
    \lambda \lesssim 5.366,
\end{equation}
as   illustrated in Fig.~\ref{shadow3}.

\subsection{Kaniadakis entropy}

Kaniadakis statistics provide a deformation of the standard Boltzmann-Gibbs 
framework and naturally arises in systems characterised by non-Gaussian 
behaviour, long-range correlations, or relativistic kinetic effects  
\cite{PhysRevE.66.056125,PhysRevE.72.036108}. In 
gravitational applications, the Kaniadakis deformation has been used to model 
entropy corrections in black-hole thermodynamics, holography, and cosmology. 

The 
Kaniadakis entropy is given by
\begin{equation}\label{Kaniadakis_entropy}
    S_{\kappa}
    = \frac{1}{\kappa}\,\sinh\!\left(\kappa\,\pi r_h^{2}\right),
\end{equation}
which smoothly reduces to the Bekenstein-Hawking law in the limit \(\kappa \to 
0\). For finite \(\kappa\), the entropy incorporates non-extensive statistical 
effects and possible quantum-gravitational contributions.
In black-hole and cosmological applications, the parameter \(\kappa\) is 
typically restricted to
\begin{equation}
    0 \leq \kappa \lesssim 0.5,
\end{equation}
ensuring a controlled perturbative expansion and thermodynamic stability. 
Positive \(\kappa\) corresponds to entropy enhancements associated with 
correlated or long-range interactions.

\subsubsection{Thermodynamic and topological classification}

\begin{figure}[ht]
\centering
\includegraphics[scale=0.45]{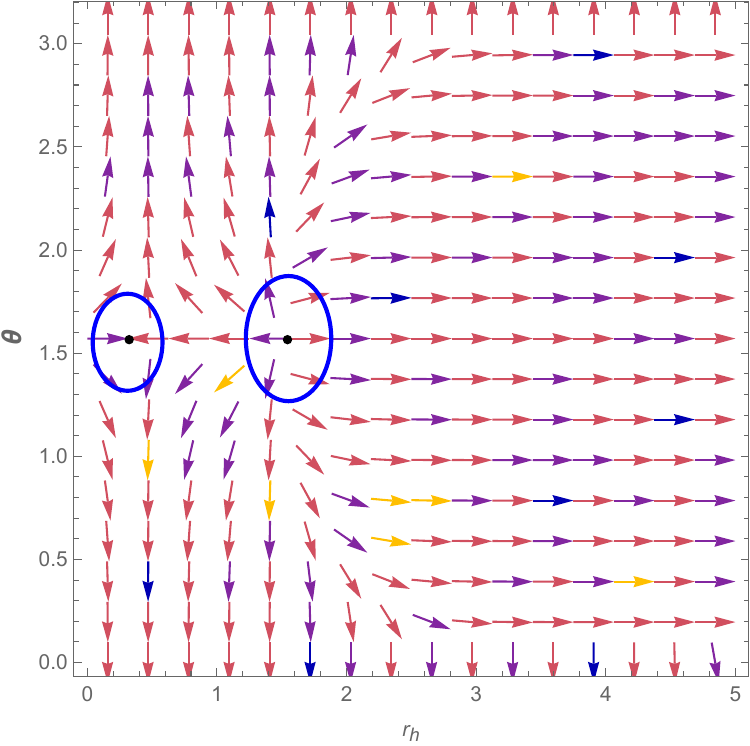}
\caption{{\it{The $r_h-\theta$ diagram in the case of Kaniadakis-modified 
black-hole entropy. As we observe,   for a fixed 
\(\tau\) 
the system admits two distinct critical radii, corresponding to 
different topological behaviours, yielding winding numbers \(w_i = +1\) and 
\(w_i = -1\), respectively.}}}
\label{Fig:Kaniadakis_R_theta}
\end{figure}

Substituting \eqref{Kaniadakis_entropy} into the off-shell free energy 
\eqref{Off_shell_free_energy_def} leads to
\begin{equation}\label{GenF_Kaniadakis}
    {\cal F}_{\kappa}(r_h)
    = \frac{1}{2}\,r_h\,\cosh\!\left(\pi\kappa r_h^{2}\right)
      - \frac{\sinh\!\left(\pi\kappa r_h^{2}\right)}{\kappa \tau}.
\end{equation}
Using the equilibrium condition \(\phi^{r_h}=0\) from \eqref{Tau_gen}, the 
auxiliary temperature parameter becomes
\begin{equation}\label{Tau_Kaniadakis}
    \tau_{\kappa}
    = \frac{4\pi r_i}
           {1 + 2\pi\kappa r_i^{2}\tanh\!\left(\pi\kappa r_i^{2}\right)}.
\end{equation}
This transcendental relation cannot be solved analytically for \(r_i\), and 
therefore the critical radius is determined numerically. As we observe from the 
   \(r_+-\theta\) portrait presented in Fig. \ref{Fig:Kaniadakis_R_theta}, 
   multiple branches may occur depending on the value of \(\tau\).
In particular, we observe that  both a stable and an unstable equilibrium 
point appear, corresponding to winding numbers of opposite sign. Their 
coexistence yields a net topological charge
\begin{equation}
    W = w_1 + w_2 = 0,
\end{equation}
indicating a topologically neutral configuration. This behaviour parallels the 
case of logarithmic entropy, in contrast to Barrow and R\'enyi deformations 
which produced a net charge \(W=-1\).

\begin{figure}[ht]
\centering
\includegraphics[scale=0.6]{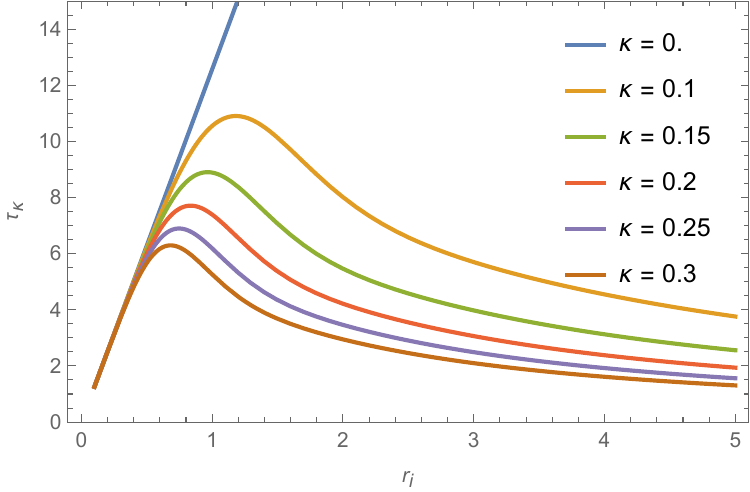}
\caption{{\it{The auxiliary temperature   parameter $\tau_{\rm log}$ versus 
the critical radius $r_i$, in the case of Kaniadakis-modified 
black-hole entropy, for different 
$\kappa$ values.   While in the Schwarzschild limit 
($\kappa=0$) there exists a single critical radius, nonzero $\kappa$ values 
produce a bifurcation in which two distinct critical radii appear at the same 
$\tau_{\rm 
log}$, indicating two topologically different branches.}}}
\label{Fig:Tau_Kaniadakis}
\end{figure}

Finally, in  Fig. \ref{Fig:Tau_Kaniadakis} we depict the auxiliary 
temperature   parameter 
$\tau_{\rm log}$ versus the critical radius $r_i$ for 
different values of 
$\kappa$. As we can see, while the Schwarzschild limit 
($\kappa=0$) yields a single critical radius, nonzero $\kappa$ values produce 
a 
bifurcation in which two distinct critical radii appear at the same $\tau_{\rm 
log}$, verifying the  two topologically different branches  corresponding to 
the two winding numbers obtained in 
Fig.~\ref{Fig:Kaniadakis_R_theta}.

\subsubsection{Photon-sphere analysis and constraints from Sgr A*}

In order to study optical signatures of the Kaniadakis deformation, we employ 
the 
modified metric function extracted in \cite{Anand:2025rjg}, namely 
\begin{equation}
    f_{\kappa}(r)
    = 1 - \frac{2M\,\text{sech}(\pi\kappa r^{2})}{r},
\end{equation}
where \(M\) is the ADM mass. Substituting this into the expression for geodesic 
curvature \eqref{geodesic_curvature} yields
\begin{equation}
    \kappa_{g}^{\kappa}(r)
    = -\frac{3M}{r^{2}\cosh(\pi\kappa r^{2})}
      - 2\pi\kappa M\,\frac{\tanh(\pi\kappa r^{2})}{\cosh(\pi\kappa r^{2})}
      + \frac{1}{r}.
\end{equation}
Solving the photon-sphere condition \(\kappa_{g}^{\kappa}(r_{ph}^{\kappa})=0\) 
perturbatively in \(\kappa\) gives
\begin{equation}
    r_{ph}^{\kappa}
    = 3M
      + \frac{81}{2}\,\pi^{2}\kappa^{2}M^{5}
      + \mathcal{O}(\kappa^{4}).
\end{equation}
\begin{figure}[ht]
\centering
\includegraphics[scale=0.6]{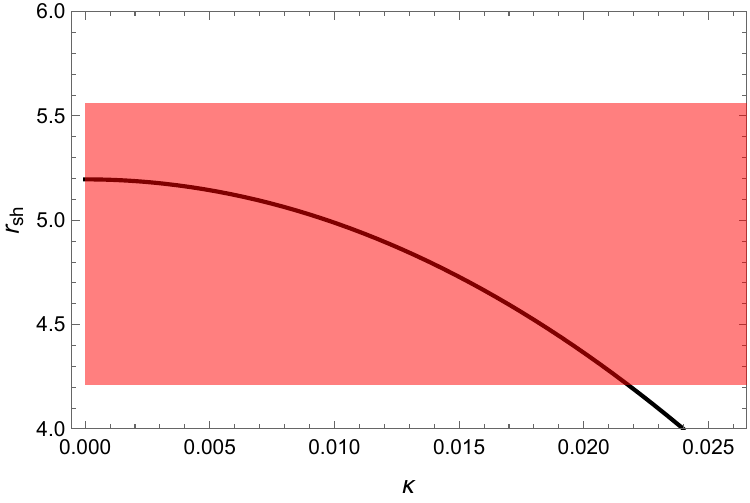}
\caption{{\it{Shadow radius for the Kaniadakis entropy corrected black hole. 
The red 
region is consistent with the EHT horizon-scale image of Sgr A at $2 \sigma 
$.}}}
\label{shadow4}
\end{figure}
The linear term in \(\kappa\) vanishes, thus the first non-trivial correction 
is 
quadratic in \(\kappa\), producing an outward shift depending on \(M^{5}\).
Additionally, using  expression \eqref{Gauss_curvature}, the Gaussian curvature 
at the photon sphere 
becomes
\begin{equation}
    \mathcal{K}^{\kappa}(r_{ph}^{\kappa})
    = -\frac{1}{27M^{2}}
      - \pi^{2}\kappa^{2}M^{2}
      + \mathcal{O}(\kappa^{3}),
\end{equation}
which is strictly negative, confirming the instability of the circular null 
orbit. 
Finally, the corresponding shadow radius is
\begin{equation}
    r_{sh}^{\kappa}
    = 3\sqrt{3}\,M
      - \frac{243}{2}\sqrt{3}\,\pi^{2}\kappa^{2}M^{5}
      + \mathcal{O}(\kappa^{3}),
\end{equation}
showing that Kaniadakis corrections produce a quadratic decrease in the shadow 
size.

\begin{table*}[t]
\centering
\caption{Summary of   the topological charge $W$, the photon-sphere 
corrections, and  observational bounds, for each modified entropy   
considered in this work.}
\begin{tabular}{lcccc}
\hline\hline
\textbf{Entropy  } 
& \textbf{ $W$} 
& \textbf{Photon-Sphere Shift $r_{ph}$} 
& \textbf{Shadow Correction $r_{sh}$} 
& \textbf{EHT Bound (Sgr A$^\ast$)} 
\\
\hline
Barrow 
& $-1$ 
& $\ \ r_{ph} = 3M - \dfrac{M}{2}\big[1+\ln(729\pi^3 M^6)\big]\Delta$
& $\ \ r_{sh} = 3\sqrt{3}M - \dfrac{3\sqrt{3}M}{2}\big[1+\ln(9\pi 
M^2)\big]\Delta$
& $\Delta \lesssim 0.08744$
\\[6pt]

R\'enyi 
& $-1$ 
& $r_{ph} = 3M + 9\pi \lambda M^3$
& $r_{sh} = 3\sqrt{3}M + 27\pi \sqrt{3}\,\lambda M^3$
& $\lambda \lesssim 0.00248$
\\[6pt]

Logarithmic 
& $0$ 
& $r_{ph} = 3M - \dfrac{5\lambda}{9\pi M}$
& $r_{sh} = 3\sqrt{3}M - \dfrac{\lambda}{\sqrt{3}\pi M}$
& $\lambda \lesssim 5.366$
\\[6pt]

Kaniadakis 
& $0$ 
& $r_{ph} = 3M + \dfrac{81}{2}\pi^2 \kappa^2 M^5$
& $r_{sh} = 3\sqrt{3}M - \dfrac{243}{2}\sqrt{3}\,\pi^2 \kappa^2 M^5$
& $\kappa \lesssim 0.02179$
\\
\hline\hline
\end{tabular}
\label{tab:summary}
\end{table*}

Lastly, using the Sgr~A$^\ast$ EHT constraint \eqref{const} with \(M=1\), the 
deformation parameter is bounded by
\begin{equation}
    \kappa \lesssim 0.02179,
\end{equation}
as illustrated in Fig.~\ref{shadow4}. This upper bound is consistent with other 
independent constraints, including Big-Bang Nucleosynthesis 
\cite{Sheykhi:2025zre} and cosmological analyses 
\cite{Hernandez-Almada:2021aiw,Hernandez-Almada:2021rjs,Yarahmadi:2024lzd,
Fang:2024yni,Luciano:2025ykr}.

\section{Conclusions}\label{Sec:Conclusions}

In this work  we investigated how deviations from the Bekenstein-Hawking 
entropy reshape both the thermodynamic and geometric properties of black holes. 
Using the recently established entropy-geometry correspondence, wherein a 
prescribed entropy functional uniquely determines the backreacted spacetime 
metric and its effective matter content, we constructed a unified framework in 
which entropy deformations are treated not as external inputs but as intrinsic 
geometric degrees of freedom. This approach enables a coherent exploration of 
how generalized entropies, motivated by quantum gravity, non-extensive 
statistics, and horizon microstructure, affect black-hole thermodynamics and 
optical properties in a self-consistent manner.

A central outcome of our analysis is the identification of distinct topological 
signatures induced by different modified entropy relations, within the 
$\phi$-mapping 
thermodynamic framework. By calculating the generalized free energy and 
examining 
its stationary points, we extracted the winding numbers associated with each 
modified entropy. As we saw, Barrow and R\'enyi deformations yield a single 
unstable 
thermodynamic sector with global charge $W=-1$, marking them as topologically 
equivalent to each other and distinct from Schwarzschild. In contrast, 
logarithmic and Kaniadakis corrections generate pairs of defects with opposite 
orientations, resulting in a net topological charge $W=0$. This cancellation 
reveals the coexistence of stable and unstable thermodynamic branches, a 
phenomenon absent in the standard Schwarzschild solution, and thus it provides 
a novel 
classification scheme for entropy-deformed black holes.

Additionally, we examined the photon-sphere geometry and shadow characteristics 
associated with each modified entropy. By evaluating the geodesic optical 
curvature, the Gaussian optical curvature, and the resulting shadow radius, we 
showed that entropy deformations generically shift the photon-sphere radius and 
induce detectable modifications of the shadow size. These effects are linear in 
the deformation parameters for Barrow, R\'enyi, and logarithmic entropies, and 
quadratic for Kaniadakis entropy, reflecting the different underlying 
statistical structures. Then, using the Event Horizon Telescope measurements 
of the 
Sgr~A$^\ast$ shadow, we derived observational upper bounds on all entropy 
parameters, thereby providing the first combined topological and optical 
constraints on these generalized entropy frameworks. For completeness, and 
reader's convenience, in Table \ref{tab:summary} we summarized the above 
results. Our findings demonstrate 
that horizon-scale imaging can serve as a direct probe of deviations from the 
Bekenstein-Hawking law. Finally, we mention that our results reveal an 
interesting correlation: modified entropy relations that introduce fractal or 
non-extensive 
statistical structure,  i.e. Barrow and R\'enyi ones, yield a net negative 
topological charge, 
whereas models with symmetric or balanced microscopic corrections, 
i.e. logarithmic and 
Kaniadakis, necessarily produce neutral configurations $W = 0$.

The present study opens several promising directions for future work. Extending 
the analysis to rotating or charged black holes would allow for a richer 
interplay between entropy deformations, frame-dragging effects, and multi-ring 
photon spheres. The entropy-geometry correspondence may also be explored in 
dynamical or cosmological settings, including gravitational collapse or 
early-universe scenarios. Additionally, combining thermodynamic topology with 
quasi-normal modes, strong-lensing observables, or accretion-disk spectra could 
yield complementary signatures of generalized entropy. With forthcoming 
improvements in Very Long Baseline Interferometry (VLBI) resolution and 
sensitivity, the framework developed here 
offers a timely avenue for confronting quantum-gravity-motivated 
modified entropies
with increasingly precise astrophysical observations.

\section*{Acknowledgements}

A.A is financially supported by the Institute's postdoctoral fellowship at IIT 
Kanpur. S.B. and
E.N.S. gratefully acknowledges  the 
contribution of 
the LISA Cosmology Working Group (CosWG), as well as support from the COST 
Actions CA21136 -  Addressing observational tensions in cosmology with 
systematics and fundamental physics (CosmoVerse)  - CA23130, Bridging 
high and low energies in search of quantum gravity (BridgeQG)  and CA21106 -  
 COSMIC WISPers in the Dark Universe: Theory, astrophysics and 
experiments (CosmicWISPers).

\bibliographystyle{utphys}
\bibliography{ref}

\end{document}